\documentclass{article}
\usepackage{frascatiphys,here,graphicx,subfigure}
\begin{document}
\title{SYSTEMATIC ERRORS OF TRANSITION FORM FACTORS
EXTRACTED BY MEANS OF LIGHT-CONE SUM RULES}
\author{Wolfgang Lucha
\\{\em Institute for High Energy Physics,
Austrian Academy of Sciences,}
\\{\em Nikolsdorfergasse 18, A-1050,
Vienna, Austria}\\
Dmitri Melikhov\\
{\em Institute for High Energy
Physics, Austrian Academy of Sciences,}\\
{\em Nikolsdorfergasse
18, A-1050, Vienna, Austria}\\
{\em and}\\{\em Nuclear Physics
Institute, Moscow State University,}\\
{\em 119992, Moscow, Russia}\\
Silvano Simula\\{\em INFN, Sezione di Roma III,}\\{\em
Via della Vasca Navale 84, I-00146, Roma, Italy}}
\maketitle
\baselineskip=11.6pt

\begin{abstract}
This talk presents results of our study of
heavy-to-light transition form factors extracted with the help of
light-cone sum rules. We employ a model with~scalar particles
interacting via massless-boson exchange and study the
heavy-to-light correlator, relevant for the extraction of the
transition form factor. We calculate this correlator in two
different ways: by making use of the Bethe--Salpeter wave function
of the light bound state and by making use of the light-cone
expansion. This allows us to calculate the full
correlator and separately the light-cone contribution to it. In this way
we show that the off-light cone contributions~are not suppressed
compared to the light-cone one by any large parameter.
Numerically, the difference between the value of the form factor
extracted from the full correlator and from the light-cone
contribution to this correlator is found to be about 20--30\% 
in a wide range of masses of the particles involved in the decay
process.
\end{abstract}
\baselineskip=14pt

\newpage
In a previous talk\cite{talk1} (see also Ref.~[2] for details) we
have shown that~the hadron parameters can be extracted from sum
rules only with some accuracy, which lies beyond the control of
the standard procedure adopted in the method of sum rules, even if
the correlator in a limited range of the Borel parameter is
known {\it precisely}. In the light-cone sum-rule analysis of
hadron form factors, the relevant correlator is not known
precisely and is obtained as an expansion near the light cone
(LC)\cite{lcsr}. This entails additional uncertainties in the
extraction~of hadron parameters, in this case, of the form factors.
This talk reports the results of our recent systematic analysis of off-light-cone 
effects in sum rules~for heavy-to-light
form factors\cite{lms_lcsr}. 

The effects are investigated in a model involving~scalar constituents. 
We consider two types of scalar ``quarks'', viz., heavy quarks $Q$
of mass~$m_Q$ and light quarks $\varphi$ of mass $m,$ and study
the weak transition of the heavy scalar ``meson'' $M_Q(Q\varphi)$
to the light ``meson'' $M(\varphi\varphi)$ induced by the weak
heavy-to-light $Q\to\varphi$ quark transition. The
analysis of this model is technically simpler but allows one to
study some essential features of the corresponding QCD case.

For calculating the correlator of interest, we need the
Bethe--Salpeter (BS) amplitude of the light meson, defined by
\begin{eqnarray}
\label{BS}
\Psi_{\rm BS}(x,p')=\langle 0|T\varphi(x)\varphi(0)|M(p')\rangle=\Psi(x^2,xp',p'^2=M^2).
\end{eqnarray}
As a function of $xp'$, this amplitude may be represented by the
Fourier integral
\begin{eqnarray}
\label{nak1} \Psi_{\rm BS}(x,p')=\int\limits_0^1{\rm d}\xi
\exp({-{\rm i}\xi p'x})K(x^2,\xi),
\end{eqnarray}
where the $\xi$-integration runs from $0$ to $1$. The kernel
$K(x^2,\xi)$ may be expanded near the light cone $x^2=0$:
\begin{eqnarray}
\label{logseries}
K(x^2,\xi)=\phi_0(\xi)+x^2\,\phi_1\left(\xi,\log(-x^2)\right)+{\rm
O}(x^4).
\end{eqnarray}
It is convenient to use the parametrization of $K(x^2,\xi)$
proposed by Nakanishi\cite{nakanishi}
\begin{eqnarray}
\label{nak2} K(x^2,\xi)=\frac{1}{(2\pi)^4{\rm
i}}\int\limits_0^\infty{\rm d}z\, G(z,\xi)\int \frac{{\rm
d}^4k'\exp({-{\rm i}k'x})}{[\,z+m^2-\xi(1-\xi)M^2-k'^2-{\rm
i}0]^3},
\end{eqnarray}
where $G(z,\xi)$ exhibits no singularities in the integration
regions in $z$ and $\xi$.~The function $G(z,\xi)$ may be obtained
as the solution of an equation obtained from~the BS equation for
$\Psi_{\rm BS}(x,p')$. The LC distribution amplitudes $\phi_i$ can
be expressed in terms of $G(z,\xi)$. For instance, the light-cone distribution amplitude reads 
\begin{eqnarray}
\label{2.6}
\phi_0(\xi)&=&\frac{1}{32\pi^2}\int\limits_0^\infty{\rm
d}z\,\frac{G(z,\xi)}{z+m^2-\xi(1-\xi)M^2}.
\end{eqnarray}
For interactions dominated by exchange of a massless boson at
small distances, the solution of the BS equation in the ladder
approximation takes the form\cite{karmanov}
\begin{eqnarray}
G(z,\xi)=\delta(z)G(\xi), \qquad G(\xi)=\xi(1-\xi)f(\xi),
\end{eqnarray}
where $f(\xi)$ is nonzero at the end-points. In this case, all
distribution amplitudes exhibit the same end-point behaviour,
namely,
\begin{eqnarray}
\label{das} \phi_0(\xi)\simeq \xi,\qquad \phi_1(\xi)\simeq
\xi,\qquad \dots.
\end{eqnarray}
Now, to extract the $M_Q\to M$ transition form factor, we analyze 
the correlator
\begin{eqnarray}
\label{correlator} \Gamma(p^2,q^2)={\rm i}\int{\rm d}^4x
\exp({{\rm i}px})
\langle 0|T\varphi(x)Q(x)Q(0)\varphi(0)|M(p')\rangle.
\end{eqnarray}
We should (i) write this correlator as a dispersion representation in
$p^2$ 
\begin{eqnarray}
\label{disp_th} \Gamma_{\rm th}(p^2,q^2)=\int\frac{{\rm d}s}{s-p^2-{\rm i}0}
\Delta_{\rm th}(s,q^2),
\end{eqnarray}
(ii) perform the Borel transform $p^2\to \mu^2$ which gives 
\begin{eqnarray}
\label{disp_th1} 
\Gamma_{\rm th}(p^2,q^2)\to 
\hat\Gamma_{\rm th}(\mu^2,q^2)=\int {{\rm d}s}\exp\left(
-s/2\mu^2\right)\Delta_{\rm th}(s,q^2),
\end{eqnarray}
and (iii) cut the correlator at an effective continuum threshold
$s=s_0$ getting
\begin{eqnarray}
\label{3.9} \hat \Gamma_{\rm th}(\mu^2,q^2,s_0)= \int{\rm d}s\,
\theta(s<s_0)\,\exp\left({-s/2\mu^2}\right)\Delta_{\rm th}(s,q^2).
\end{eqnarray}
The form factor is related to the cut correlator by
\begin{eqnarray}
\label{srff} f_{M_Q}\,F_{M_Q\to
M}(q^2)=\exp\left(M_Q^2/2\mu^2\right) \hat\Gamma_{\rm
th}\left(\mu^2,q^2,s_0(\mu^2,q^2)\right),
\end{eqnarray}
where $f_{M_Q}$ is the decay constant of the heavy meson $M_Q$ and
$s_0(\mu^2,q^2)$ is an effective continuum threshold, 
dependent on both $q^2$ and $\mu^2$.

For large $m_Q$ and for $q^2\ll m_Q^2$, up to terms
power-suppressed by $1/m_Q^2$, the correlator reads\cite{lms_lcsr}
\begin{eqnarray}
\label{3.5} 
\Gamma_{\rm th}(p^2,q^2)=\int\frac{{\rm d}^4k\,{\rm
d}^4x}{(2\pi)^4} 
e^{{\rm i}x(p-k)}
\frac{1}{m_Q^2-k^2-{\rm i}0} \langle
0|T\varphi(x)\varphi(0)|M(p')\rangle.
\end{eqnarray}
In order to calculate this correlator, we may proceed along two
different lines. 

\noindent 
I. Express the correlator in terms of the BS
amplitude $\Psi_{\rm BS}$ in momentum~space:
\begin{eqnarray}
\Gamma_{\rm th}(p^2,q^2)=\frac1{(2\pi)^4}\int{\rm d}^4k
\frac{\Psi_{\rm BS}(k,p')}{m_Q^2-(p-k)^2-{\rm i}0}.
\end{eqnarray}
It is then straightforward to calculate $\Delta_{\rm th}(s,q^2)$
in terms of the kernel $G(z,\xi)$. The corresponding explicit
expression for $\Gamma_{\rm th}$ may be found in Ref.~[4]. 

\noindent II. Use the
light-cone expansion of $\Psi_{\rm BS}(x,p')$:
\begin{eqnarray}
\label{3.22} 
\Gamma_{\rm th}(p^2,q^2)=\int\frac{{\rm d}^4k\,{\rm
d}^4x }{(2\pi)^4}
%\frac{\exp\left({\rm i}x(p-k)\right)}{m_Q^2-k^2-{\rm i}0}
e^{{\rm i}x(p-k)}\frac{1}{m_Q^2-k^2-{\rm i}0}
\sum_{n=0}^{\infty}(x^2)^n\int\limits_0^1{\rm d}\xi
%\exp(-{\rm i}p'x\xi)
e^{-{\rm i}p'x\xi}
\phi_n(\xi),
\end{eqnarray}
with the functions $\phi_i(\xi)$ related to $G(z,\xi)$.

Let us introduce the following quantities: the binding energy of the heavy
hadron $\varepsilon_Q$ by $M_Q=m_Q+\varepsilon_Q$; a new Borel
parameter $\beta$ by $\mu^2=m_Q\beta$; a new effective continuum
threshold $\delta$ by $s_0=(m_Q+\delta)^2$, such that $\varepsilon
< \delta < \beta$. The parameters $\varepsilon$, $\delta$, and $\beta$ 
remain finite in the limit $m_Q\to\infty$.
Hereafter, the light-meson mass is set equal to zero: $M=0.$ We
consider the case $q^2=0,$ and suppress the argument $q^2$ in
the correlators.

The uncut Borel image ({\it not} related to the form factor of interest) reads 
\begin{eqnarray}
\label{3.24a}
e^{\frac{M_Q^2}{2m_Q\beta}}\hat \Gamma_{\rm th}(\beta)=\int\limits_0^1 \frac{d\xi}{1-\xi}
\left[
\phi_0(\xi)-\frac{1}{\beta^2}\frac{\phi_1(\xi)}{(1-\xi)^2}+\cdots \right]
\exp\left(-\frac{m_Q \xi}{2\beta(1-\xi)}\right).  
\end{eqnarray}
For large $m_Q$, the integral is saturated by region of small $\xi=O(\beta/m_Q)$. 

The cut Borel image, i.e. the l.h.s. of (\ref{srff}) which yields the heavy-to-light 
form factor, takes the form [one should be careful with the surface terms when applying 
the cut in the dispersion representation, see details in ref.[4]]:
\begin{eqnarray}
\label{cutlc1}
e^{\frac{M_Q^2}{2m_Q\beta}}
\hat \Gamma_{\rm th}(\beta,\delta)=\int\limits_0^{\xi_0} \frac{d\xi}{1-\xi}
\left[
\phi_0(\xi)-\frac{\phi_1(\xi)}{\beta^2(1-\xi)^2}+\cdots \right]
\exp\left(-\frac{m_Q \xi}{2\beta(1-\xi)}\right)
\nonumber\\	
-4 \exp\left(\frac{\varepsilon_Q-\delta}{\beta}\right)
\left[
\frac{\phi_1(\xi_0)}{m_Q^2}
+\frac{\phi_1(\xi_0)}{2m_Q\beta}+\frac{\phi'_1(\xi_0)}{m_Q^2}
\right]+\cdots, 
\end{eqnarray}
where $\xi_0=2\delta/m_Q$ and $\cdots$ stand for the contributions
of terms corresponding~to $n\ge 2$ and of terms power-suppressed
for large $m_Q$.

Let us now address an important question: Are the off-LC contributions 
(which represent one of the higher-twist effects)  
suppressed compared to the light-cone contribution?

In the {\it uncut\/} correlator, the off-LC terms are suppressed
by powers of the parameter $1/\beta$ (but remain of the same order
in $1/m_Q$ as the LC contribution).

For the {\it cut\/} correlator, however, the situation is quite
different because of the presence of surface terms. We may consider the following 
cases: $\delta, m \ll \beta$, while $m_Q\to \infty$ and $\delta, m \ll m_Q$, while $\beta\to
\infty$. Due to the end-point behaviour of the distribution amplitudes (\ref{das}), 
in both cases the contributions of the terms $n=0, 1, \ldots$ have the same order.
Therefore we conclude that {\it for the realistic case of
interactions dominated by massless-boson exchange at short
distances, the off-LC contributions are not suppressed compared to
the LC contribution by any~large~parameter}.

\begin{figure}[ht]
\begin{center}
\begin{tabular}{cc}
\includegraphics[width=5.55cm]{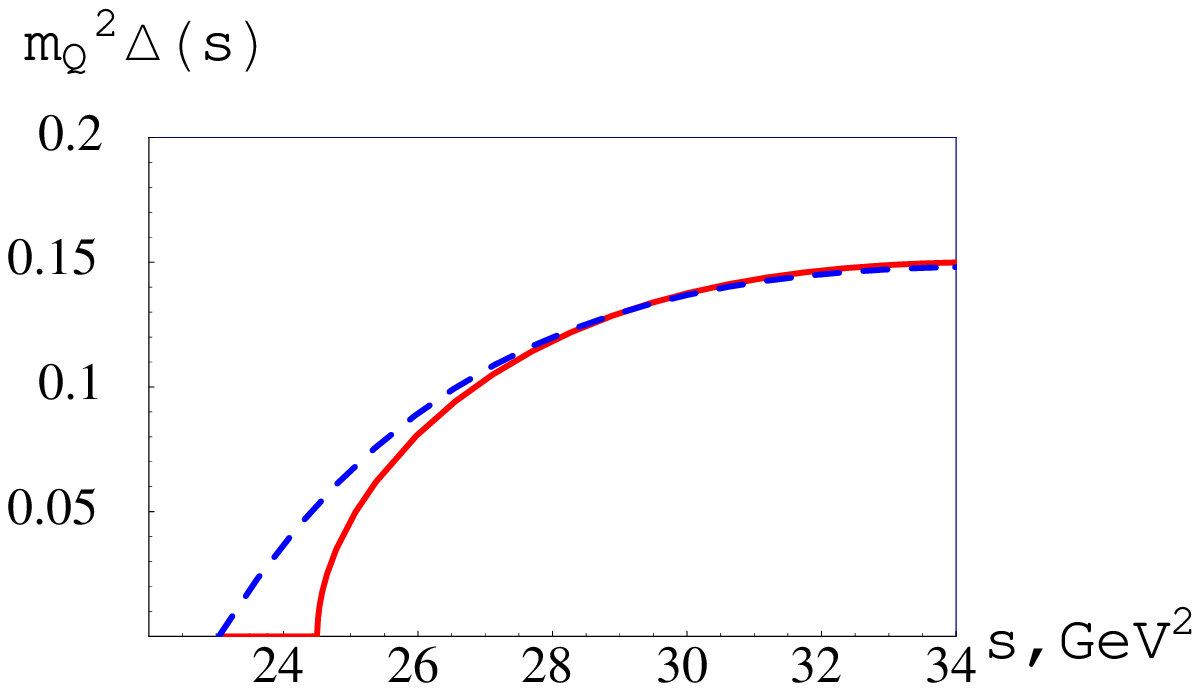}&
\includegraphics[width=5.55cm]{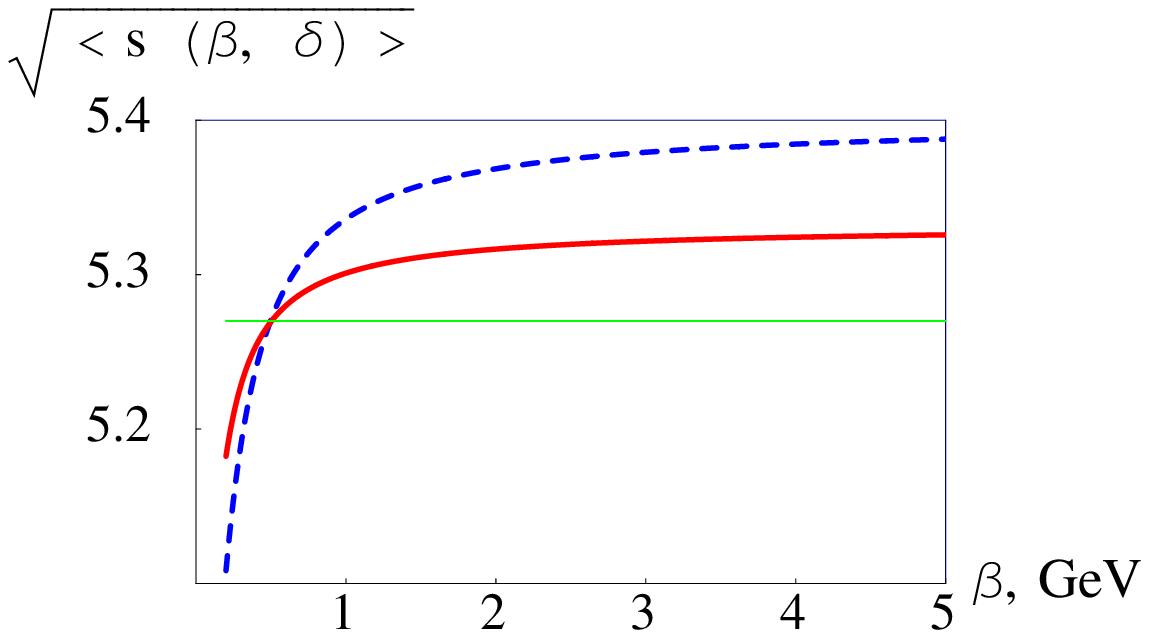}\\
\includegraphics[width=5.55cm]{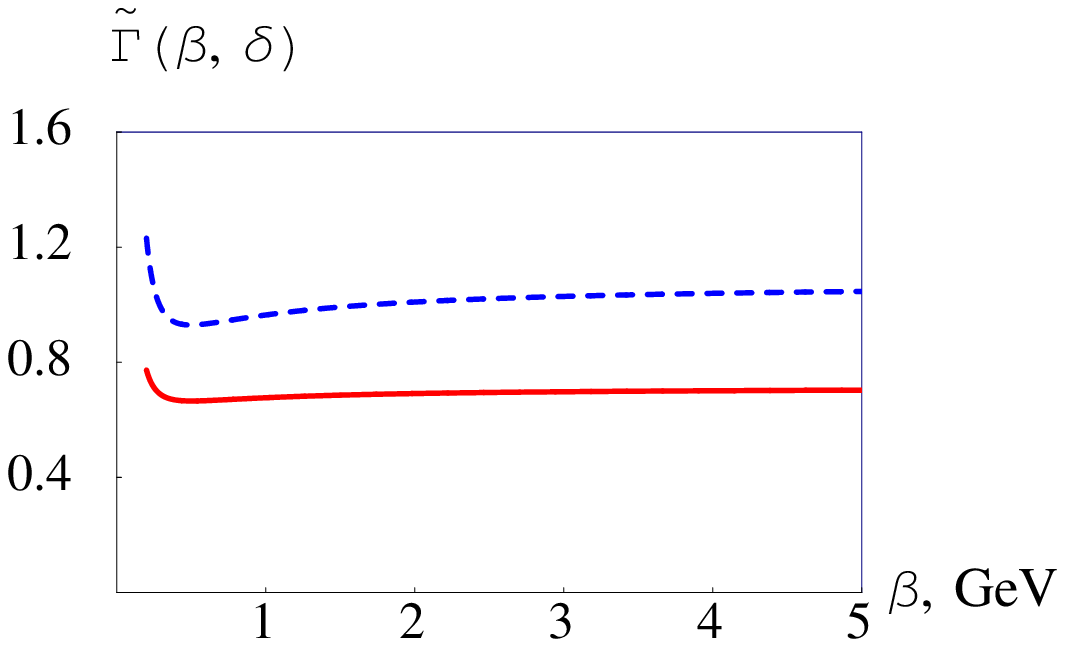}&
\includegraphics[width=5.55cm]{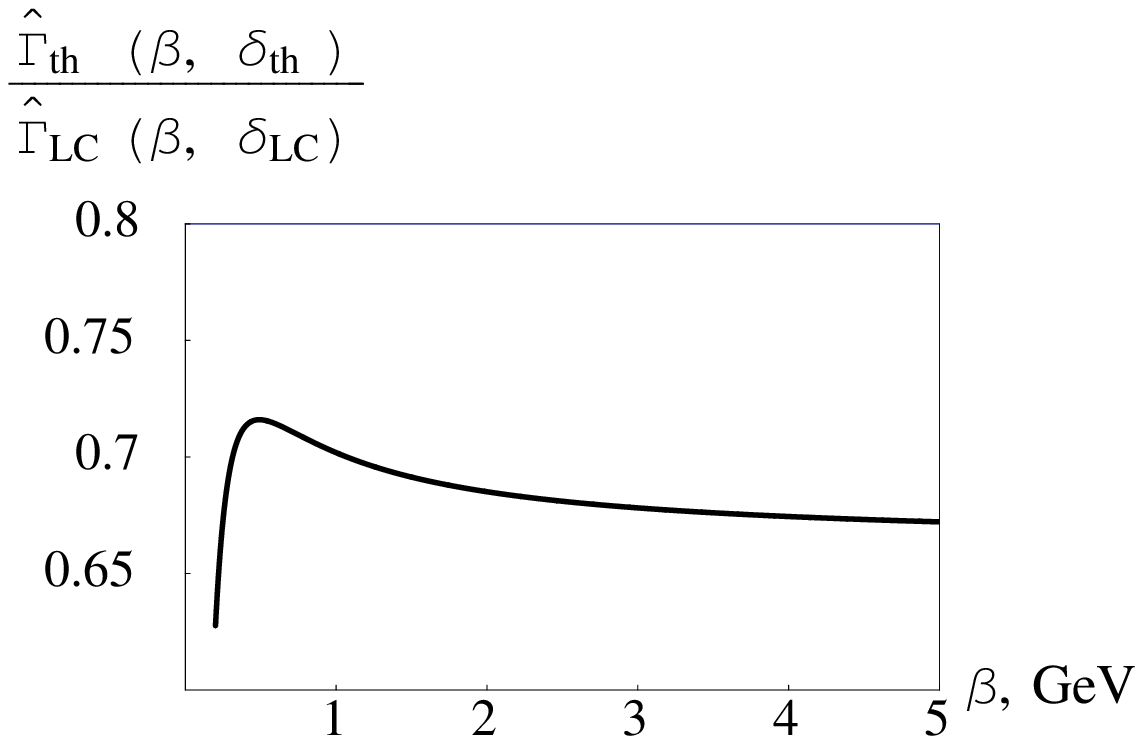}\\
\end{tabular}
\end{center}
\caption{\label{Fig:3}
Plots for the parameters corresponding to
beauty-meson decay $m_Q=4.8$ GeV, $m=150$ MeV, $\delta_{\rm
LC}=0.96$ GeV, and $\delta_{\rm th}=0.79$ GeV. {\it Upper left
panel\/}: Spectral densities $m_Q^2\Delta_{\rm th}(s)$ (solid red
line) and $m_Q^2\Delta_{\rm LC}(s)$ (dashed blue line). {\it Upper
right panel\/}: $\sqrt{\langle s \rangle_{\rm th}}$ (solid red
line) and $\sqrt{\langle s \rangle_{\rm LC}}$ vs.\ $\beta$ (dashed
blue line). The horizontal (green) line locates $M_Q=5.27$ GeV.
{\it Lower left panel\/}:
$\widetilde\Gamma(\beta,\delta)=m_Q^2\exp\left(M_Q^2/(2\mu^2)\right)
\hat\Gamma(\mu^2,s_0) $ vs.\ $\beta$: $\widetilde\Gamma_{\rm
th}(\beta,\delta_{\rm th})$ (solid red line) and
$\widetilde\Gamma_{\rm LC}(\beta,\delta_{\rm LC})$ (dashed blue
line). {\it Lower right panel\/}: The ratio $\hat\Gamma_{\rm
th}(\beta,\delta_{\rm th})/\hat\Gamma_{\rm LC}(\beta,\delta_{\rm
LC})$ vs.\ $\beta$.}\end{figure}

Next, we give numerical estimates. Fig.~\ref{Fig:3} shows results
for beauty-meson decay, with $M_Q=5.27$ GeV$, m_Q=4.8$ GeV, and
$m=150$ MeV. The discussion of the relevant parameter values and
further examples may be found in Ref.~[4].

Hereafter, the $n=0$ contribution to the correlator in Eq.~(\ref{3.22}) 
is referred to as the light-cone correlator; 
$\Delta_{\rm LC}(s)$ is the corresponding spectral density. 

Taking into account that the end-point region is essential for the
transition form factors, we can without loss of generality take the kernel of the form
$G(z,\xi)=m^2 \delta(z)\xi(1-\xi)$. It is then straightforward to
calculate the spectral densities $\Delta_{\rm th}$ and $\Delta_{\rm
LC}$ [cf.~Fig.~\ref{Fig:3}]. It is important that the thresholds
in $\Delta_{\rm th}$ and $\Delta_{\rm LC}$ do not coincide: in the
light-cone spectral density the threshold is $m_Q^2$ whereas in
the full spectral density~it is $(m_Q+m)^2$. The region near the
threshold provides the main contribution~to the cut
Borel-transformed correlator. The mismatch of the 
thresholds is responsible for the nonvanishing of the
off-light-cone effects in the cut correlator.

The effective continuum threshold $\delta$ is the quantity which
determines~to~a great extent the values of hadron observables
extracted from the sum~rule\cite{lms_sr}. We fix $\delta$ by a
standard procedure: we require that, for both LC and full spectral
densities,
\begin{eqnarray}
\label{delta2} \langle s(\beta,\delta) \rangle=M_Q^2.  
\end{eqnarray}
This equation may be used as the definition of the implicit function
$\delta(\beta)$. We, however, proceed in a different way: we do not consider the $\beta$-dependent
$\delta_{\rm th}$ and $\delta_{\rm LC}$, but determine constant
values $\delta_{\rm th}$ and $\delta_{\rm LC}$ such that relation
(\ref{delta2}) is satisfied for some specific value of $\beta$.
Here, $\delta$ is fixed from
\begin{eqnarray}
\label{delta3}
\sqrt{\langle s \rangle_{\rm LC}}=\sqrt{\langle s \rangle_{\rm th}}=M_Q
\end{eqnarray}
for $\beta=0.5$ GeV; this gives $\delta_{\rm LC}=0.96$ GeV and $\delta_{\rm th}=0.79$ GeV.

As can be seen from the plots, the light-cone contribution to the
correlator considerably exceeds the full correlator. Obviously,
the difference between these two quantities is just the
contribution of the off-LC terms in the LC expansion of the
correlator. This difference is to a large extent of pure
``kinematical''~origin, related to the mismatch between the
thresholds in $\Delta_{\rm th}$ and $\Delta_{\rm LC}$. 

\newpage
The main results of the present analysis may be summarized as follows:

\vspace{.1cm}
1. The difference between the cut full correlator and the LC
contribution to the latter is always nonvanishing, since the 
off-LC contributions are not suppressed by any large parameter
compared to the LC one. {\it In
heavy-to-light decays, there exists no rigorous theoretical limit
in which the cut LC correlator coincides with the cut full
correlator}.

2. The light-cone contribution provides {\it numerically} the bulk of
the cut full correlator, the contribution of the off-LC terms being 
always negative. Thus, the light-cone correlator
systematically {\it overestimates\/} the full correlator, the
difference at small $q^2$ being $20\div 30$\%. 

3. The Borel curves for the full and the LC correlators turn out
to be of similar shapes. Such a similarity of the Borel curves 
implies that the systematic difference between the
correlators cannot be diminished~by any relevant choice of the
criterion for extracting the heavy-to-light form factor.

Finally, let us point out the following: Although the model
discussed here differs, in many aspects, from QCD, it mimics
correctly those features which~are essential for the effects
discussed. Therefore, many of the results derived~in~this work
hold also for QCD. In particular, the following relationship
between the light-cone and the full correlators for large values
of $m_Q$ and $\mu$ is valid in QCD:
\begin{eqnarray}
\label{result_QCD} \frac{\hat \Gamma_{\rm
th}(\mu^2,q^2=0,\delta)}{ \hat \Gamma_{\rm
LC}(\mu^2,q^2=0,\delta)}= 1-{\rm
O}\left(\frac{\Lambda_{\rm QCD}}{\delta}\right).
\end{eqnarray}
For numerical estimates, we used parameter values relevant for $B$
and $D$ decays. We therefore believe that also the numerical
estimates for off-LC effects (one~of the higher-twist effects)
obtained in this work provide a realistic estimate for
higher-twist effects in QCD.

Thus, our analysis suggests a sizeable contribution to
heavy-to-light correlators, related to higher-twist effects in
QCD. This contribution is hard to control in the method of
light-cone sum rules because higher-twist distribution amplitudes
are not known with sufficient accuracy. Therefore, one might
expect sizeable errors in the heavy-to-light form factors, related
to higher-twist effects. [These errors arise in addition to the
systematic errors related to the procedure of extracting hadron
observables from a correlator discussed in our first
talk\cite{talk1}]. The effect is larger for decays of heavy mesons
containing the strange quark,~i.e., of $B_s$ and $D_s$, than for
the decays of $B$ and $D$ mesons.

The off-LC and other higher-twist effects in weak decays of heavy
mesons in QCD deserve a detailed investigation: for the method of
light-cone sum rules the corresponding distribution amplitudes are
``external'' objects and should be provided by other
nonperturbative methods. In particular, the combination~of
light-cone sum rules with approaches based on the constituent
quark picture\cite{melikhov}, which successfully describe
heavy-meson decays, might be fruitful. Moreover,~it seems
promising to apply different versions of QCD sum rules to
transition form factors\cite{braguta}; this may be helpful in
understanding the genuine uncertainties of~the form factors
extracted from the light-cone sum rules.

\vspace{3ex}\noindent{\bf Acknowledgements.} D.~M. would like to
thank the Austrian Science Fund (FWF) for support under project
P17692.


\begin{thebibliography}{30}
\bibitem{talk1}W.~Lucha, D.~Melikhov, and S.~Simula, 
``Systematic errors of bound-state parameters extracted 
by means of SVZ sum rules'', arXiv:0712.0177.  
\bibitem{lms_sr}W.~Lucha, D.~Melikhov, and S.~Simula, Phys.~Rev.~D
{\bf 76}, 036002 (2007); Phys.~Lett.~B {\bf 657}, 148 (2007);
W.~Lucha and D.~Melikhov, Phys.~Rev.~D {\bf 73}, 054009 (2006);
Phys.~Atom.~Nucl.~{\bf 70}, 891 (2007).
\bibitem{lcsr}I.~I.~Balitsky, V.~M.~Braun, and A.~V.~Kolesnichenko,
Nucl.~Phys.~B {\bf 312}, 509 (1989); V.~M.~Braun and I.~Filyanov,
Z.~Phys.~C {\bf 44}, 157 (1989); V.~I.\ Chernyak and
I.~R.~Zhitnitsky, Nucl.~Phys.~B {\bf 345}, 137 (1990); P.~Ball and
V.~M.~Braun, Phys.~Rev.~D {\bf 58}, 094016 (1998).
\bibitem{lms_lcsr}W.~Lucha, D.~Melikhov, and S.~Simula, Phys.~Rev.~D {\bf
75}, 096002 (2007).
\bibitem{nakanishi}N.~Nakanishi, Phys.~Rev.~{\bf 130}, 1230 (1963).
\bibitem{karmanov}V.~A.~Karmanov and J.~Carbonell, Eur.~Phys.~J.~A
{\bf 27}, 1 (2006).
\bibitem{melikhov}D.~Melikhov, Phys.~Rev.~D {\bf 53}, 2460 (1996);
Phys.~Rev.~D {\bf 56}, 7089 (1997); Eur.~Phys.~J.~direct {\bf C4},
2 (2002) [hep-ph/0110087]; D.~Melikhov and S.\ Simula,
Eur.~Phys.~J.~C {\bf 37}, 437 (2004);
D.~Melikhov and B.~Stech, Phys.~Rev.~D~{\bf 62}, 014006 (2000).
\bibitem{braguta}V.~Braguta, W.~Lucha, and D.~Melikhov,
arXiv:0710.5461 [hep-ph].
\end{thebibliography}
\end{document}